# Magnetic vortex oscillator driven by dc spin-polarized current

V. S. Pribiag, I. N. Krivorotov[1], G. D. Fuchs, P. M. Braganca, O. Ozatay, J. C. Sankey, D. C. Ralph, R. A. Buhrman

*Cornell University, Ithaca, New York, 14853 USA*

Abstract:

Transfer of angular momentum from a spin-polarized current to a ferromagnet provides an efficient means to control the dynamics of nanomagnets. A peculiar consequence of this spin-torque, the ability to induce persistent oscillations of a nanomagnet by applying a dc current, has previously been reported only for spatially uniform nanomagnets. Here we demonstrate that a quintessentially nonuniform magnetic structure, a magnetic vortex, isolated within a nanoscale spin valve structure, can be excited into persistent microwave-frequency oscillations by a spin-polarized dc current. Comparison to micromagnetic simulations leads to identification of the oscillations with a precession of the vortex core. The oscillations, which can be obtained in essentially zero magnetic field, exhibit linewidths that can be narrower than 300 kHz, making these highly compact spin-torque vortex oscillator devices potential candidates for microwave signal-processing applications, and a powerful new tool for fundamental studies of vortex dynamics in magnetic nanostructures.

---

[1] *Current address: University of California, Irvine, California, 92697 USA.*

A spin-polarized electron current can apply a torque on the local magnetization of a ferromagnet. This spin-transfer effect[1,2] provides a new method for manipulating magnetic systems at the nanoscale without the application of magnetic fields and is expected to lead to future data storage and information processing applications[3]. Experiments have demonstrated that spin-torque can be used to induce current-controlled hysteretic switching, as well as to drive persistent microwave dynamics in spin-valve devices[3,4,5,6,7,8,9,10,11,12]. While it is known that spin-torque switching of a magnetic element can sometimes occur via non-uniform magnetic states[13], a central remaining question is whether spin-torque can be used to efficiently excite steady-state magnetization oscillations in strongly non-uniform magnetic configurations in a manner suitable for fundamental investigations of nanomagnetic dynamics and improved device performance. A relatively simple type of non-uniform magnetic structure is a magnetic vortex, the lowest-energy configuration of magnetic structures just above the single-domain length scale[14]. Previous studies, typically performed on single-layer permalloy (Py) structures, focused on the transient or resonant response of a magnetic vortex to an applied magnetic field and identified the lowest excitation mode of a vortex as a gyrotropic precession of the core[15,16,17,18]. It has also been demonstrated that the vortex core polarization can be efficiently switched by short radio-frequency magnetic field pulses[19]. Recently, the spin-transfer effect has been used to drive a magnetic vortex into resonant precession by means of an alternating current incident on a single Py dot[20]. Here we report by means of direct frequency-domain measurements that a dc spin-polarized current can drive highly coherent gigahertz-frequency steady-state oscillations of the magnetic vortex in a nanoscale magnetic device. The high sensitivity of our technique means that fine changes in the details of the vortex oscillations, such as due to device or material inhomogeneities, can be readily detected.

The samples studied have a spin valve geometry consisting of a thick (60 nm) Py (Py = $Ni_{81}Fe_{19}$) ferromagnetic layer and a thin (5 nm) Py ferromagnetic layer separated by a 40-nm-thick Cu spacer. The thickness of the 60 nm Py layer is chosen to be above the threshold thickness necessary for the nucleation of a magnetic vortex[21]. Electron-beam lithography and ion milling were used to define and etch the spin valves resulting in pillar-shaped devices with 160 nm x 75 nm elliptical cross-section (Fig. 1D, left inset). The sample is dc current-biased along the pillar axis through Cu electrodes. Relative oscillations of the magnetizations of the two Py layers produce a time-varying voltage via the giant magneto-resistance (GMR) effect. This oscillatory output is then detected using a 30 Hz - 50 GHz spectrum analyzer. The measurements were performed at room temperature for static magnetic fields applied either perpendicular to the ellipse plane ($H_\perp$) or in-plane, parallel to the ellipse major axis ($H_\parallel$). We observe coherent microwave signals only when electrons flow from the thin Py layer towards the thick Py layer, which we define as the positive current polarity.

Figure 1A shows the dependence of the differential resistance (d$V$/d$I$) of one of the nanopillar devices on $H_\parallel$. The differential resistance curve has the typical features associated with vortex nucleation and annihilation (cf. Fig. 3 (a) in Ref. 22). Micromagnetic simulations based on the OOMMF package[23] confirm the existence of a vortex in the thick layer, while the thin layer magnetization is quasi-uniform due to the layer's reduced thickness, which makes the vortex state energetically unfavorable[21]. As $H_\parallel$ increases from zero the differential resistance decreases gradually as the vortex core approaches the device boundary. For $|H_\parallel| > \sim 650$ Oe the vortex is annihilated and both the thick and the thin layers are in quasi-uniform magnetization states with magnetic moments aligned with the field and the device resistance at its minimum. As $H_\parallel$ is reduced, near ±200 Oe the thin layer moment reorients due to the interlayer dipole field interaction[7], becoming antiparallel to the thick layer moment. The additional switching near $H_\parallel = \pm 100$ Oe corresponds to vortex nucleation, which reduces the GMR from its maximum value in the uniform, antiparallel configuration. Discrete steps in $dV/dI$ are observed for both out-of-plane (not shown) and in-plane applied fields, resulting from intermittent pinning of vortex due to material defects and device shape imperfections[24,25]. Due to the thermally activated nature of the vortex nucleation, in some scans, such as the one shown in Fig. 1a, the parallel configuration is preferred near $H_\parallel = 0$ when ramping down the field from negative values. A scan performed immediately afterwards indicates that the vortex is nucleated again near 200 Oe.

To study the spin-torque excitation of vortex oscillations the dc bias current $I$ was varied while keeping $H$, either $H_\parallel$ or $H_\perp$, fixed. As expected, microwave dynamics are observed only for values of $H_\parallel$ between the positive and negative vortex annihilation fields. Figure 1b shows typical frequency-domain measurements of the GMR signal for $H_\perp = 1600$ Oe, as a function of $I$ for sample 1, measurements that are indicative of the excitation of a single strong mode of persistent high-frequency magnetization dynamics in the structure. While we usually observe only one dominant microwave mode under the $H$ and $I$ conditions of interest here, depending upon the bias and sample we sometimes find multiple modes. In general, the power in the second harmonic signal is less than 10% of that in the fundamental, consistent with a nearly sinusoidal oscillation in the time domain.

The linewidth is typically narrow (60 MHz to <0.3 MHz), indicating that the oscillation is highly coherent, however we observe considerable variation with field and current bias conditions. For sample 1, as $I$ is increased for $H_\perp = 1600$ Oe, the full width at half-maximum ($\Delta f$) decreases, while the power density increases rapidly, reaching a maximum at $I = 11$ mA (Fig. 1b). The oscillation frequency ($f$) increases nearly linearly with $I$ (Fig. 1b, inset) at a rate of ~30 MHz/mA. Figure 1c shows the current dependence of the microwave oscillations for the same sample but for a lower field, $H_\parallel = 540$ Oe. Both $\Delta f$ and the power density show less variation with $I$ than for the $H_\perp$ case, but the frequency $f$ again exhibits an approximately linear dependence on $I$ (Fig. 1c, inset) with a similar rate of change of ~25 MHz/mA. For both in-plane and out-of-plane fields, the oscillations are characterized by $f/\Delta f$ factors that can be $> 10^3$. Figure 1d shows a high-resolution plot of a peak with $\Delta f = 2.8 \times 10^2$ kHz and $f/\Delta f = 4.0 \times 10^3$, observed for $H_\parallel = 480$ Oe and $I = 9.0$ mA. The largest $f/\Delta f$ factors previously observed in a spin-torque oscillator, on the order of $10^4$, were obtained in applied magnetic fields that were one order of magnitude larger than for the data in Fig. 1d[26]. As might be expected from a vortex system, we have observed coherent oscillations ($f/\Delta f > 10^2$) for $H_\parallel$ as small as ~6 Oe (Fig. 1d, right inset).

To gain more in-depth understanding of the vortex oscillations we compared the data to micromagnetic simulations of the Landau-Lifshitz-Gilbert (LLG) equation with an additional spin-transfer term. Figure 2a shows the equilibrium configuration of the vortex in the thick ferromagnetic layer for $H_\perp$ = 200 Oe, in the absence of spin-torque. The initial magnetic configuration of the thin ferromagnetic layer (not shown) is nearly uniform, with the in-plane component of the magnetic moment pointing in the −x direction.

The simulation indicates that the vortex enters an oscillatory regime (Fig. 2b) in the presence of a spin-polarized current $I$ with positive polarity, with the core precessing in a larger trajectory at the top surface of the thick layer than at the bottom. This motion occurs as a consequence of the transfer of spin-angular-momentum from the incident current to the local moments near the top surface of the thick layer. The resulting torque on the vortex drives it away from its equilibrium position into a trajectory that is further determined by magnetostatic restoring forces[17]. The spin-polarization of the incident current has a spatial distribution that replicates the magnetic structure of the thin ferromagnetic layer, which itself undergoes periodic oscillations from its quasi-uniform, in-plane equilibrium state. These thin layer oscillations have the basic character of periodically flexing the thin layer magnetization into configurations with C-state and S-state-like components, depending upon the details of the simulation conditions, as the vortex precesses.

The simulations predict an increase in the oscillation frequency with $I$ at a rate of ~30 MHz/mA, consistent with the experimentally observed rate of 20-50 MHz/mA, a result which can be attributed in part to the stronger confinement of the vortex motion by the Oersted field with larger $I$. The simulations also show that the motion of the vortex is asymmetric about the equilibrium position of the vortex. This symmetry breaking is due to the quasi-uniform spin polarization incident from the thin layer and typically results in an elongated vortex trajectory that is rotated with respect to the ellipse axes (Fig. 2c). As the vortex follows this elongated trajectory it undergoes periodic distortions from the ideal cylindrically symmetric shape, which in turn cause oscillatory changes in the magnetoresistance of the nanopillar device, the origin of the microwave voltage. For $I$ greater than ~20 mA, the simulations indicate that the vortex begins to deform significantly due to the large spin-torque, while the GMR signal becomes more chaotic, in agreement with experimental observations on device 1 that the peak broadens and eventually disappears for $I$ greater than ~15 mA.

For negative $I$ bias the electrons that impinge on the thin layer have acquired a vortex-shaped spin polarization distribution in passing previously through the thick layer. The micromagnetic simulations show that the resulting spin-torque acting on the thin layer causes its magnetization to curl into a vortex. As the magnetization of the thin layer becomes aligned with the spin-polarization the spin-transfer torques on both the thick and thin layers vanish, leading to a stationary steady state, consistent with no coherent microwave oscillations being observed experimentally for negative $I$. When the current is turned off in the simulation, the thin layer magnetization unwinds to the quasi-uniform state due to the shape anisotropy field.

The measured vortex oscillation frequency $f$ shows qualitatively different behaviors for $H_\perp$ and $H_\parallel$, with $f$ increasing with $H$ in the first case (Fig. 3) and decreasing in the second case (Fig. 4), a trend that is reproduced by the simulations. Increasing $H_\perp$ reduces the non-uniformity of the magnetic vortex along the z axis, bringing it closer to an ideal vortex, while the opposite occurs for in-plane applied fields. Thus, we attribute the field-dependence of $f$ to this straightening (deformation) of the vortex for $H_\perp$ ($H_\parallel$)[18]. Uncertainties in the device dimensions introduce a maximum relative uncertainty of ~20% in the values of $f$ obtained from the simulations.

While the $H_\perp$ data for sample 2 shown in Fig. 3 exhibits a monotonic frequency dependence with a single oscillation mode, the sample 1 data for $H_\parallel$ (Fig. 4) includes a series of small discontinuities between roughly linear regions, as well as several regions where two or three peaks coexist in the time-average spectrum. We attribute these jumps in the oscillation frequency to abrupt changes in the details of the vortex's confining magnetostatic potential due to magnetic defects in the thick Py layer, or to abrupt changes in the polarization distribution of the incident spin-torque current due to defects in the thin Py layer. The presence of such defects, which could be intrinsic, or due to shape anomalies or antiferromagnetic surface oxides[27], is confirmed by the existence of irregularities in the GMR field scan (Fig. 1a) near the field values where the $f$ vs. $H_\parallel$ data show discontinuities. As a result of these defects, in certain bias regimes the vortex can undergo jumps between slightly different, metastable trajectories, corresponding to transitions between local minima of the confining potential. Evidence for such metastable trajectories can be seen in the lower inset in Fig. 4 where for $H_\parallel$ = 350 Oe, either two or three distinct microwave peaks, with approximately 60 MHz separations, are observed in the time-averaged spectrum. In studies of the transient oscillation of single vortices in micrometer scale Py disks, the vortex gyrotropic frequency has been observed to fluctuate by as much as a factor of two as changes in the dc field moves the vortex equilibrium position between pinning sites as little as 10 nm apart[28]. Here we are observing finer changes in the average frequency of persistent vortex motion due to what must be only small changes in the confining potential when averaged over a trajectory that, according to the simulations, displaces the core ~ 20 nm or more from its equilibrium position.

The spin-torque-driven vortex oscillations can exhibit significantly narrower linewidths than spin-transfer oscillations in "vortex-free" magnetic nanopillar spin valves whose minimum linewidths vary between 550 MHz and ~10 MHz at room temperature[3,29]. We can qualitatively attribute this difference to the larger magnetic volume involved in the vortex oscillation and the relatively weak dependence of the oscillation frequency on magnetic field. The former reduces the amplitude of the random Langevin magnetic field associated with the thermal fluctuations, while the latter reduces the effect of this fluctuation field on the oscillation linewidth[30]. Conversely, the narrow linewidth results in this vortex oscillator being a sensitive indicator of magnetic defects in the nanostructure. If some of the magnetic defects that contribute to the non-ideal confining potential are sufficiently uncoupled to the rest of the magnetic system that they thermally fluctuate in a quasi-independent manner, they will collectively modulate the precessional frequency and broaden the oscillator linewidth. As the field and current bias conditions change, the vortex core samples different regions of the nanoscale device, and hence different defect ensembles. Improvements in materials and device fabrication should reduce these variations with bias and further narrow the oscillator linewidth beyond what we have demonstrated here in this initial experiment. Nevertheless, the oscillator is already sufficiently robust that relatively low frequency electromagnetic signals can be coherently detected, as demonstrated in the upper inset of Fig. 4, which shows the mixing response to an external 5 MHz magnetic field $H_{FM}$ applied in plane.

Our direct frequency-domain measurements demonstrate that a dc spin-polarized current can efficiently drive persistent microwave-frequency oscillations of a strongly non-uniform nanomagnetic state. This extends the range of known motions that can be excited by spin-polarized currents and provides a new avenue for studying the properties of magnetic vortices. Compared to the dynamics of uniform nanomagnets, the spin-torque-driven precession of a magnetic vortex exhibits linewidths that can be orders of magnitude narrower. In addition, the sensitivity of the vortex oscillator linewidth and frequency to local magnetic defects in the nanostructure makes it a powerful new nanoscale probe of magnetic thin film materials. This sensitivity also points to possible device improvements that could lead to even narrower linewidths. The high coherence and the ability to electrically tune the oscillation frequency suggest that this new spin-torque vortex oscillator effect could prove useful for devising nanoscale microwave oscillators and for more complex signal processing. The demonstrated ability to operate such a device in near-zero applied field could lead to easier integration of such an oscillator with current semiconductor technology.

**METHODS**

The micromagnetic simulations integrate the LLG equation with a spin-torque term of the form described in reference 31, using $\Lambda = 2$ for the torque asymmetry parameter. The material parameters are typical for Py: the damping parameter $\alpha = 0.014$, the exchange constant $A = 1.3$ μerg/cm, the saturation magnetization $M_s = 800$ emu/cm$^3$ for the thick layer, $M_s = 600$ emu/cm$^3$ for the thin layer (based on SQUID magnetometry measurements) and spin polarization[32] $P = 0.37$. The volume is discretized into 5 x 5 x 2.5 nm$^3$ elements. The simulation includes the magnetic coupling between the two ferromagnetic layers, as well as the Oersted field due to $I$. Temperature effects are not taken into account. The spin polarization of electrons transmitted to the second magnetic layer is mapped from the magnetization vector field of the first magnetic layer along the electron flow direction. We treat spins classically and use the simplifying assumption that the spin component anti-parallel to the local magnetization is fully reflected at the interface between the spacer and the second ferromagnetic layer, thus exerting a torque on the first layer. The micromagnetic simulations provide useful insight into the nature of the spin-torque-driven vortex oscillator, however we emphasize that they may not accurately describe all features of the real system due to simplified modeling of the device geometry and spin-transfer torque, and due to the absence of defects in the simulations. In particular, the linewidths obtained from simulation are limited by the finite simulation time and do not represent the effects of thermal broadening mechanisms. The oscillation amplitude in the simulations typically decays, albeit quite slowly, as a function of time, suggesting that our simulations, while fairly closely replicating the frequency of the spin-torque-driven vortex precession, and its field and current dependence, do not capture the full complexity of the spin-torque and magnetic field interlayer couplings. To our knowledge, there are no published results on micromagnetic simulations that consider the full dynamic coupling between the two ferromagnetic layers in a spin valve with spin-transfer torque. Our results suggest this coupling might be of considerable importance in some geometries, such as the vortex oscillator presented in this paper.

**ACKNOWLEDGEMENTS:**

We thank Paul Crowell for useful discussions and materials and Michael Donahue for helpful guidance on the OOMMF simulations. This research was supported in part by the National Science Center through the NSEC program support for the Center for Nanoscale Systems, by ARO - DAAD19-01-1-0541 and by the Office of Naval Research/MURI program. Additional support was provided by NSF through use of the facilities of the Cornell Nanoscale Facility - NNIN and the facilities of the Cornell MRSEC.

**Figure Legends:**

**Figure 1 GMR and microwave data for sample 1. a,** Differential resistance for $I = 0$ as a function of $H_\parallel$. The included lead resistance in (a) and (b) is ~12 Ω. The black arrows indicate the field scanning direction. **b,** Microwave spectra as a function of dc current bias for $H_\perp$ = 1600 Oe. *FWHM* (circles) and *f* (triangles) as a function of *I* (inset). **c,** Microwave spectra as a function of dc current bias for $H_\parallel$ = 540 Oe. *FWHM* (circles) and *f* (triangles) as a function of *I* (inset). The curves in (b) and (c) are offset by 0.02 nW/GHz along the vertical axis for clarity. **d,** Microwave peak with $\Delta f = 2.8 \times 10^2$ kHz and $f/\Delta f = 4.0 \times 10^3$ for $H_\parallel$ = 480 Oe and $I$ = 9.0 mA. The continuous line is a Lorentzian fit to the data. Sample layout (left inset). Microwave peak for $I$ = 10 mA and $H_\parallel$ = ~6 Oe (right inset).

**Figure 2 Micromagnetic simulation for $I$ = 6.6 mA and $H_\perp$ = 200 Oe. a,** Initial magnetic configuration of the thick magnetic layer showing in-plane components (arrows) and z-component (color shading). The cross-section is taken at the top surface of the thick ferromagnetic layer. **b,** Power density spectrum of the y-component of the thick layer magnetization showing a peak at *f* = 1.25 GHz. **c,** Magnetic configuration of the thick layer, showing the in-plane components (arrows) of the layer's upper and lower surfaces and the vortex core (red), at 0.2 ns intervals. The maximum core displacement for these values of *I* and *H* is ~20 nm. The magnetization of the thin layer (not shown) is quasi-uniform, with its in-plane components undergoing small-amplitude oscillations from the –x direction, as described in the text.

**Figure 3 Dependence of microwave frequencies on $H_\perp$ for sample 2.** The blue dots show the sample 2 data as a function of $H_\perp$ for $I$ = 6.6 mA. The black squares indicate the results from micromagnetic simulations on a 160 nm x 75 nm ellipse.

**Figure 4 Dependence of microwave frequencies on $H_\parallel$ for sample 1.** The green dots show the sample 1 data as a function of $H_\parallel$, for $I$ = 12 mA. The black squares indicate results from simulations on a 160 nm x 75 nm ellipse. Power density plot for $H_\parallel$ = 350 Oe (lower inset). Demonstration of frequency modulation of a microwave peak, showing first-order side-bands (upper inset). The dc applied magnetic field is $H_\parallel$ = 550 Oe and the dc current is $I$ = 9 mA. The frequency modulation is induced by an oscillating magnetic field $H_{FM}$ with a frequency $f_{FM}$ = 5 MHz, applied along the minor axis of the ellipse.

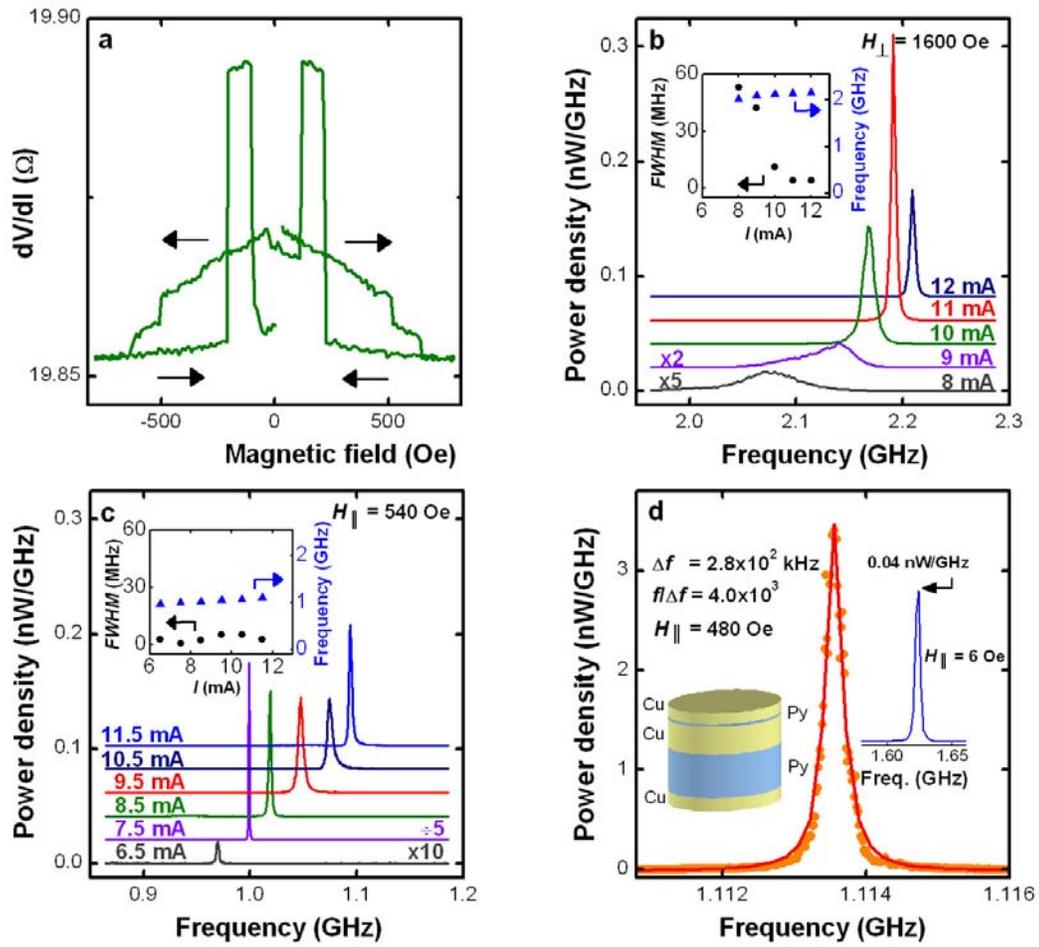

**Figure 1**

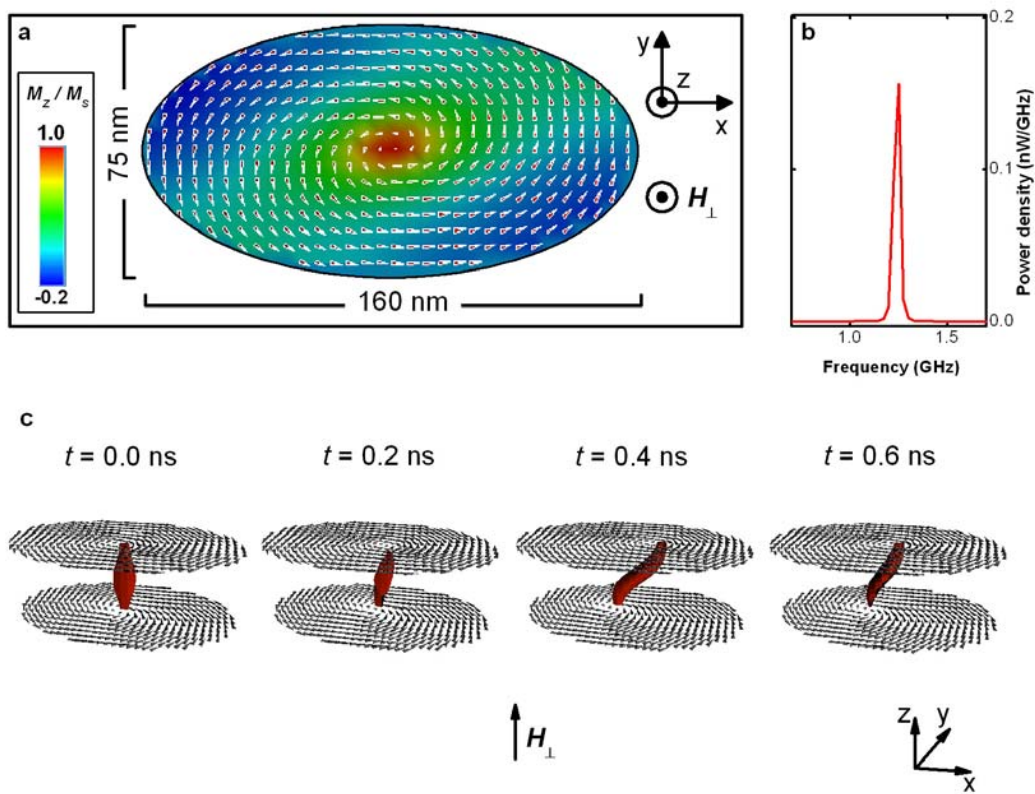

**Figure 2**

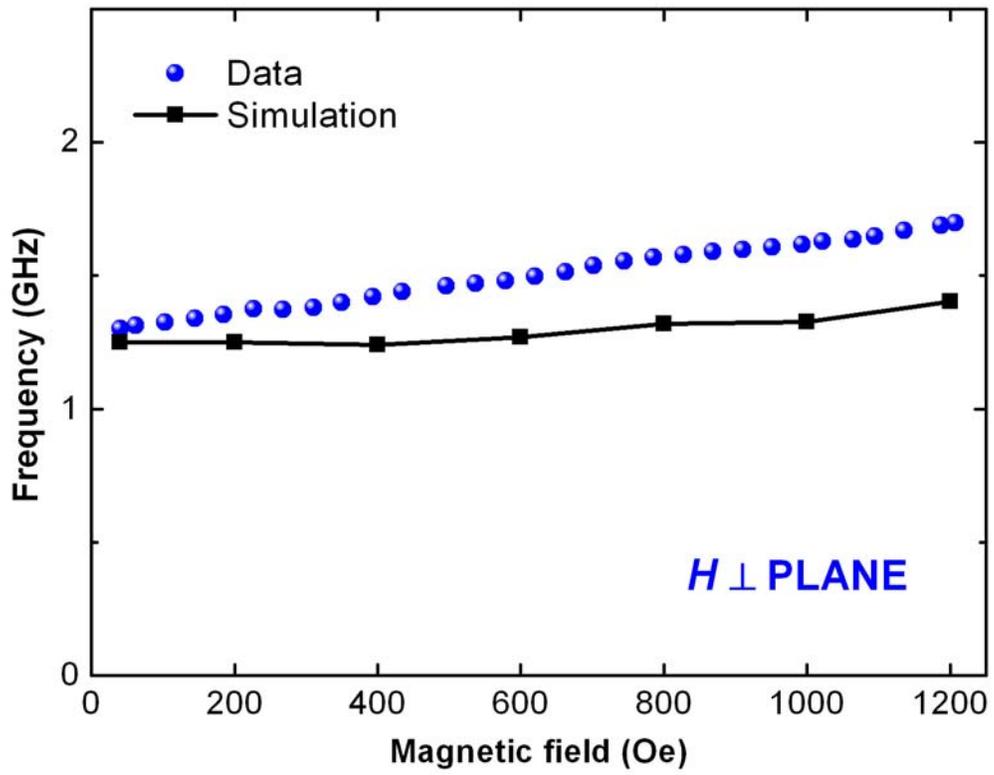

**Figure 3**

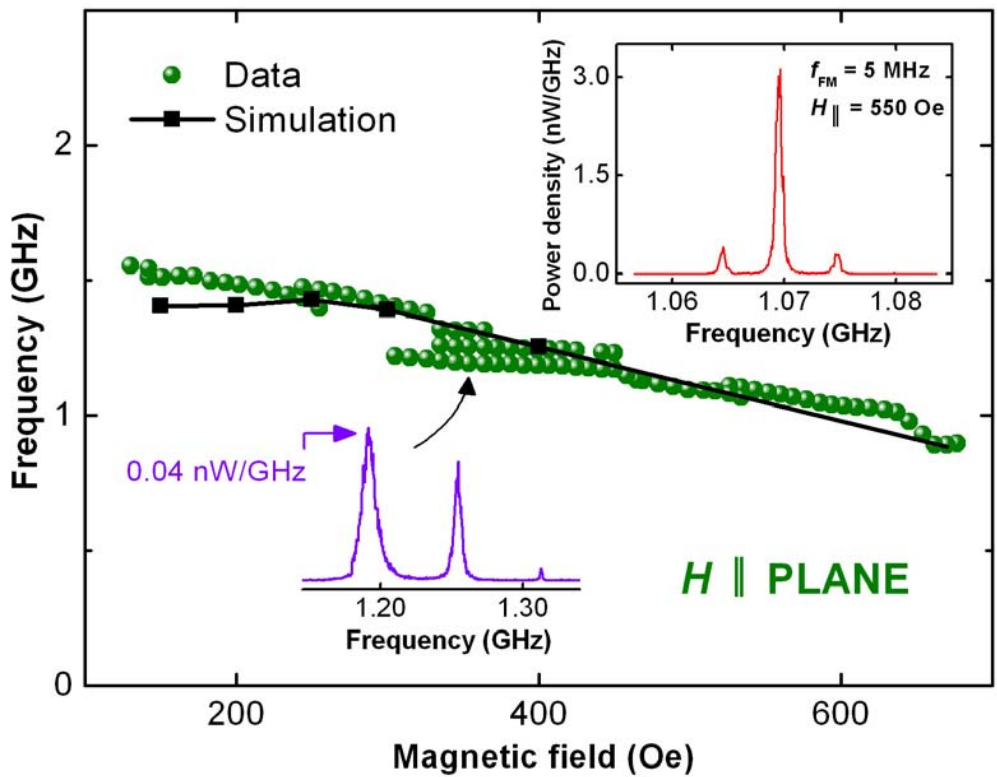

**Figure 4**